\documentclass[twocolumn,english,aps,floatfix,superscriptaddress,amsfonts,amssymb,superscriptaddress,prb]{revtex4}
\usepackage{color}

\usepackage{graphicx}
\usepackage{amsmath}
\usepackage{latexsym}
\usepackage{epstopdf}
\usepackage{amsmath,amssymb}
\usepackage{amssymb,amsmath}
\usepackage{multirow}
\usepackage{mathtools}
\usepackage{bbold}
\usepackage[dvipsnames]{xcolor}
\newcommand{\beq}{\begin{equation}}
\newcommand{\eeq}{\end{equation}}

\usepackage{tikz}
\usepackage{xcolor}
\usetikzlibrary{patterns}
\allowdisplaybreaks

\begin{document}

\title{Nonequilibrium spectral moment sum rules of the Holstein-Hubbard model}

\author{Khadijeh Najafi}
\affiliation{Department of Physics, Harvard University, Cambridge, Massachusetts 02138, USA}
\affiliation{IBM Quantum, IBM T.J. Watson Research Center, Yorktown Heights, NY 10598 USA}
\affiliation{Department of Physics, Georgetown University, 37th and O Sts. NW, Washington, DC 20057, USA}

\author{J.~Alexander~Jacoby}
\affiliation{Department of Physics, Box 1843, 182 Hope St., Brown University, Providence, RI 02912, USA}

\author{R.~D.~Nesselrodt}
\affiliation{Department of Physics, Georgetown University, 37th and O Sts. NW, Washington, DC 20057, USA}

\author{J.~K.~Freericks}
\affiliation{Department of Physics, Georgetown University, 37th and O Sts. NW, Washington, DC 20057, USA}
\date{\today{}}

\begin{abstract}

We derive a general procedure for evaluating the  {\rm $n$th} derivative of a time-dependent operator in the Heisenberg representation and employ this approach to calculate the zeroth to third spectral moment sum rules of the retarded electronic Green's function and self-energy for a system described by the Holstein-Hubbard model allowing for arbitrary spatial and time variation of all parameters (including spatially homogeneous electric fields and parameter quenches).  For a translationally invariant (but time-dependent) Hamiltonian, we also provide sum rules in momentum space. The sum rules can be applied to various different phenomena like time-resolved angle-resolved photoemission spectroscopy and benchmarking the accuracy of numerical many-body calculations. This work also corrects some errors found in earlier work on simpler models.

\end{abstract}
\pacs{}
\maketitle


\section{Introduction}
Recent developments in pump-probe spectroscopy have elucidated some nonequilibrium properties of a large variety of strongly correlated systems with coupled electrons, phonons, and spin degrees of freedom to femtosecond time scales\cite{{Abreu,Coslovich,Schmitt,Stojchevska,Hellmann,Giannetti,Smallwood}}. For example, pump-probe techniques have been applied to study high $T_{c}$ cuprates, which exhibit archetypal strong electron-electron and electron-phonon couplings\cite{Graf,Rettig,Rameau,tatiana}. Despite these inroads, understanding the nonequilibrium dynamics of electron-phonon interactions and their interplay with electron-electron interactions at large remains elusive and is among the most intriguing problems in condensed matter physics.

First, we briefly review the ``pump-probe'' technique. The ``pump'' part refers to an ultrastrong and ultrashort electric field pulse, which can be used to selectively excite either electrons or phonons. The resulting nonequilibrium state can subsequently be explored by a ``probe,'' which is a weaker pulse that measures the response of the system after a delay. Among the different materials that have been studied by pump-probe spectroscopy, high temperature superconductors have been central, in part because the role played by electron-phonon interactions in these materials is still not well-understood. For example, Zhang {\it et al.}\cite{Zhang}  investigated the ultrafast response of the self-energy of a high-temperature superconductor in both the normal and superconducting states. Such studies are valuable as the most direct evidence of an electron-phonon coupling in the cuprates (or, more generally, an electron-boson coupling) is the universal electron self-energy renormalization, which manifests itself as a kink in the photoemission spectrum that occurs below the Fermi energy precisely at the coupled phonon energy\cite{Lanzara1}.

In equilibrium, the strength of the kink is directly related to the strength of the electron-phonon coupling. Whether this phenomenon is related to high-temperature superconductivity still remains unclear. One intriguing result from pump-probe experiments\cite{Zhang} is that the kink softens when in the superconducting state, even with a relatively weak pump. Is the pump dynamically reducing the electron-phonon coupling in superconducting states? Sum rules are one way to answer this question. A numerical study \cite{Kemper}, shows that the kink softens when the system is pumped, even if there is no dynamic reduction of the electron-phonon coupling as determined by the zeroth moment of the retarded self-energy as a function of time. Hence, kink softening alone is insufficient to determine whether there is any dynamic reduction of the electron-phonon coupling, even though it is routinely used to determine the strength of electron-phonon coupling in equilibrium photoemission studies.

Here, we explain how to use exact sum rules to investigate the effect of a pump on a system with both electron-electron and electron-phonon interactions using a Holstein-Hubbard model. A particularly advantageous feature of this model is its relative simplicity while still capturing the essential physics of electron-electron couplings and electron-phonon couplings\cite{Freericks0,Bauer1,Werner1,Tezuka,Koller}. The sum rules can also be purposed to benchmark the accuracy of numerical approaches. Indeed, the approach was developed to calculate the first two moments of the spectral function in order to estimate the accuracy of Monte-Carlo solutions of the Hubbard model in two-dimensions\cite{White}. Since then, the applications of sum rules have extended to a variety of strongly correlated systems in equilbrium and nonequibrium both for homogeneous and inhomogeneous cases\cite{Freericks1,Freericks2,Freericks3,Freericks4}. Sum rules for the retarded Green's function through second order for the Holstein model\cite{kornilovitch} and the zeroth-order self-energy sum rule for the Holstein-Hubbard model\cite{gunnarsson} have also appeared in equilibrium. Preliminary work has already found the lowest-order sum rules in the nonequilibrium Holstein model\cite{Freericks5}. Here, we focus on the full Holstein-Hubbard model and derive the nonequilibrium spectral moment sum rules through third order.

The remainder of the paper is organized as follows: In Sec. ${\rm II}$, we derive an identity for calculating the {\rm $n$th} derivative of a generic time dependent operator in the Heisenberg representation, which 
can be used to derive a general formula for the {\rm $n$th} moment of a spectral function in the case when the Hamiltonian is time-dependent. In Sec. ${\rm III}$, we introduce the Holstein-Hubbard model and we derive the exact sum rules for the spectral function of the retarded Green's function up to the third moment. Translational invariance of the system is not needed for these calculations, and we allow the parameters of the model to be spatially inhomogenous. In Sec. ${\rm IV}$, we derive the corresponding spectral moment sum rules for the retarded self-energy. For translationally invariant systems, we transform all the moments to momentum space in Sec. ${\rm V}$. To ensure that the position space results are well founded, we compare them to relevant analytic and numerical results in the atomic limit and against previous results in section ${\rm VI}$. A summary, with comments and conclusions, is provided in Sec. ${\rm VII}$.

\section{Formalism for the {\rm $n$th} spectral moment of the electronic Green's function}
\begin{figure}[hthp!]
 \centering
  \includegraphics[width=0.45\textwidth]{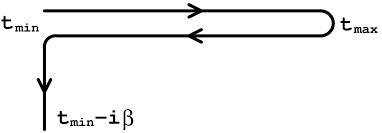}
\caption{Kadanoff-Baym-Keldysh contour with an initial thermal state at $t=t_{min}$ that has a temperature $T=1/\beta$.}
\label{Keldysh}
\end{figure}

In this section, we derive a general formula for the {\rm $n$th} moment of the non-equilibrium retarded Green's function, defined as follows:
\begin{eqnarray}\label{GR}\
G_{ij\sigma}^{R}(t,t')=- i\theta(t-t')\langle\{c_{i\sigma}(t),c_{j\sigma}^{\dagger}(t')\} \rangle,
\end{eqnarray}
where $\theta(t)$ is the unit step (or ``Heaviside'') function, $\langle O \rangle={\rm Tr}\,[\exp(-\beta \mathcal{H}_{eq})O]/\mathcal{Z}$, $\mathcal{Z}={\rm Tr}\exp(-\beta\mathcal{H}_{eq})$ and the curly bracket denotes the anticommutator ($\left\{A,B\right\} = AB + BA$) between operators. The operator $c^\dagger_{i\sigma}$ ($c_{i\sigma}$) creates (destroys) a fermion at lattice site $i$ with spin $\sigma$. These operators satisfy the canonical anticommutation relations: $\{c_{i\sigma},c_{j\sigma'}\}=0$, $\{c^\dagger_{i\sigma},c^\dagger_{j\sigma'}\}=0$ and $\{c_{i\sigma},c^\dagger_{j\sigma'}\}=\delta_{ij}\delta_{\sigma\sigma'}$. The fermionic operators are written in 
the Heisenberg representation $c_{i\sigma}(t)=U^{\dagger}(t,t_{min})\,c_{i\sigma}\,U(t,t_{min})$, where the evolution operator
satisfies the Schr\"odinger equation, $idU(t,t_{min})/dt=\mathcal{H}_S(t)U(t,t_{min})$, and the $S$ subscript denotes the Schr\"odinger representation for the Hamiltonian. The time evolution operator is 
\begin{equation}
U(t,t_{min})=\mathcal{T}_t \exp\left [ -i \int_{t_{min}}^t d\bar t \mathcal{H}_S(\bar t)\right ]
\end{equation}
where $\mathcal{T}_t$ is the time-ordering operator which moves later times to the left and $t_{min}$ is the earliest time considered in the calculation. Next, we express the retarded Green's function as
\begin{eqnarray}\label{GR2}\
&~&G_{ij\sigma}^{R}(t,t')=- i\theta(t-t')\\
&\times&\langle U^\dagger(t_{max},t_{min})U(t_{max},t)c_{i\sigma}U(t,t')c_{j\sigma}^{\dagger}U(t',t_{min}) \rangle
\nonumber\\
&-&i\theta(t-t')\nonumber\\
&\times&\langle U^\dagger(t',t_{min})c^\dagger_{j\sigma}U^\dagger(t_{max},t')U(t_{max},t)c_{i\sigma}U(t,t_{min})\rangle.
\nonumber
\end{eqnarray}
The two times in the argument of the Green's function lie on the Kadanoff-Baym-Keldysh contour, which starts from $t_{min}$ and runs in the positive 
direction until $t_{max}$ (because of the $U$ term) and then runs back
to $t_{min}$ in the opposite direction (because of the $U^\dagger$ term) and finally goes to $t_{min}-i\beta$ parallel to the imaginary axis where $\beta=1/T$ (because of the $\exp[-\beta{\mathcal H}_S(t{\to}-\infty)]$ term in the thermal average), as illustrated in Fig.~\ref{Keldysh}. In the first operator average, both $t$ and $t'$ are on the upper branch of the contour with $t$ later than $t'$, while in the second average, $t$ is on the upper branch and $t'$ is on the lower branch. In these formulas, we used the semigroup identity, $U(t_1,t_2)U(t_2,t_3)=U(t_1,t_3)$, and unitarity, $U^\dagger(t,t')U(t,t') = U(t',t)U(t,t')=\mathbb{1}$.

It is then conventional and convenient to make use of so-called Wigner coordinates: the average time is $T_{ave}=(t+t^{\prime})/2$ 
and the relative time is $t_{rel}=t-t'$. By performing the Fourier transform with respect to the relative time, $t_{rel}$, we can find the frequency-dependent retarded Green's function for each average time,
\begin{eqnarray}\label{G_omega}\
&~&G_{ij\sigma}^{R}(T_{ave},\omega)=\nonumber\\
&~&\int_{0}^{\infty}dt{_{rel}}\,e^{i\omega t_{rel}}G_{ij\sigma}^{R}(T_{ave}+\frac{t_{rel}}{2},\,T_{ave}-\frac{t_{rel}}{2}).
\end{eqnarray}

The {\rm $n$th} spectral moment in real space is then defined from the many-body density of states as
\begin{eqnarray}\label{mu_t1}\
\mu_{ij\sigma}^{Rn}(T_{ave})=-\frac{1}{\pi}\int_{-\infty}^{\infty}d\omega\,\omega^{n} {\rm Im}\, G_{ij\sigma}^{R}(T_{ave},\omega).
\end{eqnarray}
from which one may rewrite the moments as derivatives with respect to relative time as follows:
\begin{eqnarray}\label{mu_t2}\
&\,&\mu_{ij\sigma}^{Rn}(T_{ave})={\rm Im}~i^{n+1}\\
&\times&\left \langle \frac{d^{n}}{dt_{rel}^{n}}\left \{c_{i\sigma}\left (T_{ave}+\frac{t_{rel}}{2}\right ),c_{j\sigma}^{\dagger}\left (T_{ave}-\frac{t_{rel}}{2}\right )\right \}\right \rangle,\nonumber
\end{eqnarray}
evaluated in the limit as the relative time approaches zero from the positive side, $t_{rel}\rightarrow 0^+$. Thus the problem of finding the {\rm $n$th} moment of the spectral function reduces to calculating the {\rm $n$th} derivative of the time-dependent anticommutator $\{c_{i\sigma}(T_{ave}+\frac{t_{rel}}{2}),c_{j\sigma}^{\dagger}(T_{ave}-\frac{t_{rel}}{2})\}$ with respect to $t_{rel}$. 
Below, we show how to calculate this derivative in the Heisenberg representation. Consider a physical system with an arbitrary time-dependent Hamiltonian denoted by $\mathcal{H}_S(t)$ in the Schr\"odinger representation. We know that in 
the Heisenberg representation, the time dependence is
encoded in the operator $A_H(t)=U^\dagger(t,t_{min})A_S(t)U(t,t_{min})$, and the Heisenberg equation of motion implies that
\begin{eqnarray}\label{EOM2}\
i\frac{dA_H(t)}{dt}&=&U^{\dagger}(t,t_{min})[A_S(t),\mathcal{H}_{S}(t)]U(t,t_{min})\nonumber\\
&+&iU^{\dagger}(t,t_{min})\frac{\partial A_S(t)}{\partial t}U(t,t_{min}).
\end{eqnarray}
Using the definition of the $n$-fold nested commutator $L_{n}A_H(t)=[...[[A_H(t),\mathcal{H}_H(t)],\mathcal{H}_H(t)]...,\mathcal{H}_H(t)]$
and $D_{n}A_H(t)=U^\dagger(t,t_{min})\partial^n A_S(t)/\partial t^{n}U(t,t_{min})$, we can rewrite the Eq.~(\ref{EOM2}) as,

\begin{eqnarray}\label{At1}\
i\frac{dA_H(t)}{dt}=L_{1}A_H(t)+iD_{1}A_H(t).
\end{eqnarray}
Thus, one can calculate higher-order moments by taking additional derivatives. However, this procedure is unwieldy for higher derivatives, so we must simplify. We start from the schematic formula
\begin{eqnarray}\label{Atn}\
i^{n}\frac{d^{n}A_H(t)}{dt^{n}}=\sum_{{\rm sequence}=1}^{2^{n}}\,(i)^{m}\,{\rm Tuple}[\{D_{1},L_{1}\},n]A_H(t),\nonumber\\
\end{eqnarray}
where ${\rm Tuple[{list},n]}$ is an $n$-tuple, which we define to be a sequence of products of elements with length $n$ (note the curly brackets are {\it not} used as anticommutators here, nor are the square brackets commutators). The sum runs over all possible orderings of the sequences of the $n$-tuple (one can think of the orderings as all possible partitions of $L_{n}$ and $D_{n}$ regarding $L_{i}D_{j}$ as distinct from $D_{j}L_{i}$). 
The index $m$ indicates the order of the derivative for each sequence and it can be obtained by summing over the number of times that operator $D_1$ appears in the sequence. Now, we can calculate higher derivatives as follows:
\begin{eqnarray}\label{At2}\
i^{2}\frac{d^2A_H(t)}{dt^{2}}&=&L_{1}L_{1}A_H(t)+iD_{1}L_{1}A_H(t)\nonumber\\
&+&iL_{1}D_{1}A_H(t)+i^{2}D_{1}D_{1}A_H(t) \nonumber \\
&=&L_{2}A_H(t)+iD_{1}L_{1}A_H(t)\nonumber\\
&+&iL_{1}D_{1}A_H(t)+i^{2}D_{2}A_H(t),
\end{eqnarray}
\begin{eqnarray}\label{At3}\
i^{3}\frac{d^3A_H(t)}{dt^{3}}&=&L_{3}A_H(t)+iL_{1}D_{1}L_{1}A_H(t)\nonumber\\
&+&iL_{2}D_{1}A_H(t)+i^{2}L_{1}D_{2}A_H(t)\nonumber\\
&+&iD_{1}L_{2}A_H(t)+i^{2}D_{2}L_{1}A_H(t)\nonumber\\
&+&i^{2}D_{1}L_{1}D_{1}A_H(t)+i^{3}D_{3}A_H(t).
\end{eqnarray}
In deriving these formulae, we have used a contraction rule wherein one combines adjacent operator pairs that are identical operators. For example, $L_{1}L_{1}$ is equal to $L_{2}$ and $D_1D_1=D_2$. Care must be taken when evaluating the mixed operator terms, because the derivatives will act on {\it both} the $A$ operator and the $\mathcal{H}$ operator in the commutator chains, implying we must use the chain rule within the nested commutators. If the number of derivatives is smaller than the number of nested commutators, the derivatives are distributed through all possible terms. For example, if there are $n$ derivatives, then we have $2^n$ terms from the chain rule. In cases where the operator $A_S$ has no explicit time dependence simplifications are apparent. Note, however, that when the Hamiltonian has time dependence in the Schr\"odinger representation, the derivative terms have to be included because, generically, the time derivative of the Hamiltonian does not commute with the Hamiltonian. We use a tilde notation to indicate an operator that does not have explicit time dependence. Performing all of these simplifications leads to the following results for the first three derivatives when $A_S$ has no explicit time dependence in the Schr\"odinger representation:
\begin{eqnarray}\label{Att1}\
i\frac{d\tilde{A}_H}{dt}=L_{1}\tilde{A}_H,
\end{eqnarray}
\begin{eqnarray}\label{Att2}\
i^{2}\frac{d^2\tilde{A}_H}{dt^{2}}=L_{2}\tilde{A}_H+iD_{1}L_{1}\tilde{A}_H,
\end{eqnarray}
\begin{eqnarray}\label{Att3}\
i^{3}\frac{d^3\tilde{A}_H}{dt^{3}}&=&L_{3}\tilde{A}_H+iL_{1}D_{1}L_{1}\tilde{A}_H+iD_{1}L_{2}\tilde{A}_H\nonumber\\
&+&i^{2}D_{2}L_{1}\tilde{A}_H.
\end{eqnarray}
While one might think that there should be no explicit time-derivatives on the right hand side 
when the operator $\tilde A_S$ has no explicit time dependence, the derivatives enter from the nested commutators because the Hamiltonian does not generally commute with its own time derivatives when it is time-dependent.

Our next step is to employ these derivative identities to calculate the spectral moment sum rules. First, we must determine what derivatives are needed for these sum rules, so, as shown in Eq.~(\ref{mu_t2}), we start by evaluating the derivatives of the anticommutator between the creation and annihilation operators, which have no explicit time dependence in the Schr\"odinger representation. The chain rule immediately yields
\begin{widetext}
\begin{eqnarray}\label{mu_tn}\
&~&i^n\frac{d^{n}}{dt_{rel}^{n}}\left \{c_{i\sigma}\left (T_{ave}+\frac{t_{rel}}{2}\right ),c_{j\sigma}^{\dagger}\left (T_{ave}-\frac{t_{rel}}{2}\right )\right \}\\
&=&{\left(\frac{i}{2}\right)^n}\sum_{k=0}^{n} (-1)^{k}\binom {n} {k}\left \{ \left [\frac{d}{dt_{rel}}\right ]^{n-k}c_{i\sigma}\left (T_{ave}+\frac{t_{rel}}{2}\right )
,\left [\frac{d}{dt_{rel}}\right ]^{k}c_{j\sigma}^{\dagger}\left (T_{ave}-\frac{t_{rel}}{2}\right )\right \},\nonumber
\end{eqnarray}
\end{widetext}
with $\binom {n} {k}=n!/[(n-k)!k!]$ the binomial coefficient. We will take the limit $t_{rel}\rightarrow 0^+$ in all of the derivatives. Next, we employ the results from Eq.~(\ref{Atn}) to determine the first four spectral moments (with all operators in the Heisenberg representation---the subscript $H$ has been suppressed for brevity):

\begin{widetext}
\begin{eqnarray}\label{cc_R0}\
\mu_{ij\sigma}^{R0}(T_{ave})={\rm Re}{\langle}\,\{c_{i\sigma}(T_{ave}),c_{j\sigma}^{\dagger}(T_{ave})\}{\rangle},
\end{eqnarray}
\begin{eqnarray}\label{cc_R1}\
\mu_{ij\sigma}^{R1}(T_{ave})=\frac{1}{2}{\rm Re}\Big{(}{{\Big{\langle}}}\Big{\{}[c_{i\sigma}(T_{ave}),\mathcal{H}(T_{ave})],c_{j\sigma}^{\dagger}(T_{ave})\Big{\}}{\Big{\rangle}}
-{\Big{\langle}}\Big{\{}c_{i\sigma}(T_{ave}),[c_{j\sigma}^{\dagger}(T_{ave}),\mathcal{H}(T_{ave})]\Big{\}}{\Big{\rangle}}\Big{)},
\end{eqnarray}

\begin{eqnarray}\label{cc_R2}\
&~&\mu_{ij\sigma}^{R2}(T_{ave})=\frac{1}{4}{\rm Re}\Big{(}{\Big{\langle}}\Big{\{}[[c_{i\sigma}(T_{ave}),\mathcal{H}(T_{ave})],\mathcal{H}(T_{ave})],c_{j\sigma}^{\dagger}(T_{ave})\Big{\}}{\Big{\rangle}}
-2{\Big{\langle}}\Big{\{}[c_{i\sigma}(T_{ave}),\mathcal{H}(T_{ave})],[c_{j\sigma}^{\dagger}(T_{ave}),\mathcal{H}(T_{ave})]\Big{\}}{\Big{\rangle}} \nonumber 
\\
&+&{\Big{\langle}}\Big{\{}c_{i\sigma}(T_{ave}),[[c_{j\sigma}^{\dagger}(T_{ave}),\mathcal{H}(T_{ave})],\mathcal{H}(T_{ave})]\Big{\}}{\Big{\rangle}}\Big{)}
+\frac{1}{4}{\rm Im}\Big{(}{\Big{\langle}}\Big{\{}[c_{i\sigma}(T_{ave}),\frac{\partial{\mathcal{H}(T_{ave})}}{\partial{T_{ave}}},c_{j\sigma}^{\dagger}(T_{ave}){]}\Big{\}}{\Big{\rangle}}\nonumber\\
&+&{\Big{\langle}}\Big{\{}c_{i\sigma}(T_{ave}),[c_{j\sigma}^{\dagger}(T_{ave}),\frac{\partial{\mathcal{H}(T_{ave})}}{\partial{T_{ave}}}]\Big{\}}{\Big{\rangle}}\Big{)}, 
\end{eqnarray}
\begin{eqnarray}\label{cc_R3}\
&~&\mu_{ij\sigma}^{R3}(T_{ave})={\rm Re}\,\frac{1}{8}\Big{(}{\Big{\langle}}\Big{\{}[[[c_{i\sigma}(T_{ave}),\mathcal{H}(T_{ave})],\mathcal{H}(T_{ave})],\mathcal{H}(T_{ave})],c_{j\sigma}^{\dagger}(T_{ave})\Big{\}}{\Big{\rangle}}\nonumber\\
&-&3{\Big{\langle}}\Big{\{}[[c_{i\sigma}(T_{ave}),\mathcal{H}(T_{ave})],\mathcal{H}(T_{ave})],[c_{j\sigma}^{\dagger}(T_{ave}),\mathcal{H}(T_{ave})]\Big{\}}{\Big{\rangle}}
+3{\Big{\langle}}\Big{\{}[c_{i\sigma}(T_{ave}),\mathcal{H}(T_{ave})],[c_{j\sigma}^{\dagger}(T_{ave}),\mathcal{H}(T_{ave})],\mathcal{H}(T_{ave})]\Big{\}}{\Big{\rangle}}\nonumber\\
&-&{\Big{\langle}}\Big{\{}c_{i\sigma}(T_{ave}),[[[c_{j\sigma}^{\dagger}(T_{ave}),\mathcal{H}(T_{ave})],\mathcal{H}(T_{ave})],\mathcal{H}(T_{ave})]\Big{\}}{\Big{\rangle}}\Big{)}\nonumber \\  
&+&{\rm Re}\,\frac{i}{8}\Big{(}{\Big{\langle}}
\Big{\{}[[c_{i\sigma}(T_{ave}),\mathcal{H}(T_{ave})],\frac{\partial{\mathcal{H}(T_{ave})}}{\partial{T_{ave}}}],c_{j\sigma}^{\dagger}(T_{ave})\Big{\}}{\Big{\rangle}}
+2{\Big{\langle}}\Big{\{}[[c_{i\sigma}(T_{ave}),\frac{\partial{\mathcal{H}(T_{ave})}}{\partial{T_{ave}}}],\mathcal{H}(T_{ave})],c_{j\sigma}^{\dagger}(T_{ave})\Big{\}}{\Big{\rangle}}
\nonumber  \\  
&-&3{\Big{\langle}}\Big{\{}[c_{i\sigma}(T_{ave})],\frac{\partial{\mathcal{H}(T_{ave})}}{\partial{T_{ave}}},[c_{j\sigma}^\dagger(T_{ave}),\mathcal{H}(T_{ave})]\Big{\}}{\Big{\rangle}}
+3{\Big{\langle}}\Big{\{}[c_{i\sigma}(T_{ave}),\mathcal{H}(T_{ave})],[c_{j\sigma}^\dagger(T_{ave}),\frac{\partial{\mathcal{H}(T_{ave})}}{\partial{T_{ave}}}]\Big{\}}{\Big{\rangle}}\nonumber \\ 
&-&2{\Big{\langle}}\Big{\{}c_{i\sigma}(T_{ave}),[[c_{j\sigma}^{\dagger}(T_{ave}),\frac{\partial{\mathcal{H}(T_{ave})}}{\partial{T_{ave}}}],\mathcal{H}(T_{ave})]\Big{\}}{\Big{\rangle}}
-{\Big{\langle}}\Big{\{}c_{i\sigma}(T_{ave}),[[c_{j\sigma}^{\dagger}(T_{ave}),\mathcal{H}(T_{ave})],\frac{\partial{\mathcal{H}(T_{ave})}}{\partial{T_{ave}}}]\Big{\}}{\Big{\rangle}}\Big{)}\nonumber \\ 
&-&{\rm Re}\,\frac{1}{8}\Big{(}{\Big{\langle}}\Big{\{}[c_{i\sigma}(T_{ave}),\frac{\partial^{2}{\mathcal{H}(T_{ave})}}{\partial{T_{ave}}^{2}}],c_{j\sigma}^{\dagger}(T_{ave})\Big{\}}{\Big{\rangle}}
-{\Big{\langle}}\Big{\{}c_{i\sigma}(T_{ave}),[c_{j\sigma}^{\dagger}(T_{ave}),\frac{\partial^{2}{\mathcal{H}(T_{ave})}}
{\partial{T_{ave}}^{2}}]\Big{\}}{\Big{\rangle}}\Big{)},
\end{eqnarray}
\end{widetext}
Clearly, the expressions become increasingly complex for larger $n$.  There is, however, a simple solution to this problem. Start with $\{L_{0}c_{i},L_{n}c_{j}^{\dagger}\}$,
\begin{eqnarray}\label{c_ac1}\
&~&\{L_{0}c_{i\sigma}(T_{ave}),L_{n}c_{j\sigma}^{\dagger}(T_{ave})\}=\nonumber\\
&~&\{L_{0}c_{i\sigma}(T_{ave}),L_{1}L_{n-1}c_{j\sigma}^{\dagger}(T_{ave})\}.
\end{eqnarray}
We now make use of a graded Jacobi identity, with which we move the commutators to the right:
\begin{equation}
    \label{GJacobi}
 \left\{X,\left[Y,Z\right]\right\} = -\left\{\left[X,Z\right],Y\right\}+\left[\left\{X,Y\right\},Z\right].
 \end{equation}
This implies that we can rearrange Eq.~(\ref{c_ac1}) to become
\begin{eqnarray}\label{c_ac2}\
&~&\{L_{0}c_{i\sigma}(T_{ave}),L_{n}c_{j\sigma}^{\dagger}(T_{ave})\}=\nonumber\\
&~&-\{L_{1}c_{i\sigma}(T_{ave}),L_{n-1}c_{j\sigma}^{\dagger}(T_{ave})\} \nonumber \\
&~&+ L_{1}\left\{c_{i\sigma},L_{n-1}c^{\dagger}_{j\sigma}\right\}.
\end{eqnarray}
This can then be rearranged to yield 
\begin{eqnarray}
    \label{canform}
   & & \left\{L_{k}c_{i\sigma},L_{n}c^{\dagger}_{j\sigma}\right\} \nonumber \\
   &=& \sum_{i=0}^{n}\left(-1\right)^{n}\binom{n}{i}L_{n-i}\left\{L_{k+i}c_{i\sigma},c^{\dagger}_{j\sigma}\right\}.
\end{eqnarray}
This identity allows us to relate all of the multiple commutator terms to a small set of similar terms; the structure of these terms makes manifest that contributions will \textit{not} cancel as we proceed from large to small $n$ (or $k$ as in Eq.~(\ref{mu_tn})), but will contribute to fewer and fewer distinct terms. Note that here we used the symbol $L_0$, which denotes the identity (meaning no commutation).
Derivative terms in higher moments can also be simplified in this fashion, but one must be careful because the Hamiltonian and its derivative do not, in general, commute. Though these results are somewhat challenging to interpret without detailed calculation, schematically one can always plug Eq.~(\ref{canform}) directly into Eq.~(\ref{mu_tn}) to get an uglier, but easier to calculate, result.

After simplifying, we find:
\begin{widetext}
\begin{eqnarray}\label{ccc_R0}\
\mu_{ij\sigma}^{R0}(T_{ave})={\rm Re}\,{\langle}\{c_{i\sigma}(T_{ave}),c_{j\sigma}^\dagger(T_{ave})\}{\rangle},
\end{eqnarray}
\begin{eqnarray}\label{ccc_R1}\
\mu_{ij\sigma}^{R1}(T_{ave})={\rm Re}\,{\Big{\langle}}\Big{\{}[c_{i\sigma}(T_{ave}),\mathcal{H}(T_{ave})],c_{j\sigma}^{\dagger}(T_{ave})\Big{\}}{\Big{\rangle}},
\end{eqnarray}
\begin{eqnarray}\label{ccc_R2}\
\mu_{ij\sigma}^{R2}(T_{ave})=&~&{\rm Re}\,{\Big{\langle}}\Big{\{}[[c_{i\sigma}(T_{ave}),\mathcal{H}(T_{ave})],\mathcal{H}(T_{ave})],c_{j\sigma}^{\dagger}(T_{ave})\Big{\}}{\Big{\rangle}} \nonumber \\
&~&-{\rm Re}\frac{3}{4} \left<\left[\left\{\left[c_{i\sigma}\left(T_{ave}\right),H\left(T_{ave}\right)\right],c^{\dagger}_{j\sigma}\left(T_{ave}\right)\right\},H\left(T_{ave}\right)\right]\right>,
\end{eqnarray}
\begin{eqnarray}\label{ccc_R3}\
\mu_{ij\sigma}^{R3}(T_{ave})= &&{\rm Re}\,{\Big{\langle}}\Big{\{}[[[c_{i\sigma}(T_{ave}),\mathcal{H}(T_{ave})],\mathcal{H}(T_{ave})],\mathcal{H}(T_{ave})],c_{j\sigma}^{\dagger}(T_{ave})\Big{\}}{\Big{\rangle}}\nonumber\\
 && -{\rm Re}\frac{3}{2}\left<\left[\left\{\left[\left[c_{i\sigma}\left(T_{ave}\right),H\left(T_{ave}\right)\right],H\left(T_{ave}\right)\right],c^{\dagger}_{j\sigma}\left(T_{ave}\right)\right\},H\left(T_{ave}\right)\right]\right>\nonumber \\
&& +{\rm Re}\frac{3}{4}\left<\left[\left[\left\{\left[c_{i\sigma}\left(T_{ave}\right),H\left(T_{ave}\right)\right],c^{\dagger}_{j\sigma}\left(T_{ave}\right)\right\},H\left(T_{ave}\right)\right],H\left(T_{ave}\right)\right]\right> \nonumber \\ 
&&+{\rm Re}\,\frac{i}{2}\Big{(}{\Big{\langle}}\Big{\{}[[c_{i\sigma}(T_{ave}),\frac{\partial{\mathcal{H}(T_{ave})}}{\partial{T_{ave}}}],\mathcal{H}(T_{ave})],c_{j\sigma}^{\dagger}(T_{ave})\Big{\}}{\Big{\rangle}}
\nonumber  \\  \nonumber \\
&&-{\Big{\langle}}\Big{\{}[[c_{i\sigma}(T_{ave}),\mathcal{H}(T_{ave})],\frac{\partial{\mathcal{H}(T_{ave})}}{\partial{T_{ave}}}],c_{j\sigma}^{\dagger}(T_{ave})\Big{\}}{\Big{\rangle}}\Big{)}\nonumber\\
&& +{\rm Re} \frac{3}{4} \left(\left<\left[\left\{\left[c_{i\sigma},H\right],c^{\dagger}_{j\sigma}\right\},\frac{\partial H}{\partial T_{ave}}\right]\right>\right)\nonumber \\
&&-{\rm Re}\,\frac{1}{4}\Big{(}{\Big{\langle}}\Big{\{}[c_{i\sigma}(T_{ave}),\frac{\partial^{2}{\mathcal{H}(T_{ave})}}{\partial{T_{ave}}^{2}}],c_{j\sigma}^{\dagger}(T_{ave})\Big{\}}{\Big{\rangle}}\Big{)}.
\end{eqnarray}

\end{widetext}
Eqs.~(\ref{ccc_R0}-\ref{ccc_R3}) are the starting points for an explicit determination of the zeroth-third spectral moments of the retarded Green's function as presented next.

\section{Formalism for the sum rules of the spectral function for the Holstein-Hubbard model}
The Holstein-Hubbard model is widely and effectively used to describe systems with both electron-phonon and electron-electron interactions\cite{Freericks0,Bauer1,Werner1,Tezuka,Koller}. The Hamiltonian for the (inhomogeneous) Holstein-Hubbard model in the Schr\"odinger representation
is given by 
\begin{eqnarray}\label{H_HH}\
\mathcal{H}_{HH}(t)=&-&\sum_{ij\sigma}t_{ij}(t)c_{i\sigma}^{\dagger}c_{j\sigma}+\sum_{i}U_{i}(t)n_{i\downarrow}n_{i\uparrow}\nonumber \\&+&\sum_{i}[g_{i}(t)x_{i}-\mu_{i}(t)](n_{i\uparrow}+n_{i\downarrow})\nonumber\\
&+&\sum_{i}\frac{1}{2m_{i}}p_{i}^{2}+\sum_{i}\frac{1}{2} \kappa_{i}x_{i}^{2}.
\end{eqnarray}
In the above Hamiltonian, the phonon coordinate and momentum are defined as follows:
\begin{eqnarray}\label{xp}
\hat{x}=\sqrt{\frac{\hbar}{2m\omega}}(a^{\dagger}+a),\hspace{0.5cm}\hat{p}=i\sqrt{\frac{m\omega}{2\hbar}}(a^{\dagger}-a)
\end{eqnarray}
where $a^{\dagger}$ and $a$ are bosonic creation and annihilation operators, while $c^{\dagger}$ and $c$ are the creation and annihilation operators for the fermionic degrees of freedom (with the lattice site and spin indices suppressed). Consequently, $n_{i\sigma}=c_{i\sigma}^{\dagger}c_{i\sigma}$ is the occupation number of electrons of spin $\sigma$ at site $i$. The electron hopping matrix $t_{ij}(t)$ is a (possibly time-dependent) Hermitian matrix and $U_{i}(t)$ is the (possibly time-dependent) on-site Hubbard repulsion. The electrons are coupled to phonons by coupling strength $g_{i}(t)$ which is parameterized by an energy per unit length (and may also be time-dependent). A local site energy $\mu_i$ is also included (it is the chemical potential if it is independent of $i$, the lattice site). 

Note that the mass, spring constant, and frequency of the phonon are not allowed to change with time.
  
This model captures the features of a variety of interesting phenomena such as the Mott transition and polaron and bipolaron formation. It also has ordered phases to describe superconductivity, charge-density-wave order, and spin-density-wave order. Dynamical mean-field theory (DMFT) has been applied to investigate the model exactly\cite{Freericks0,Freericks6,Werner2,Hewson,Hewson2}.
 
 Now we apply Eqs.~(\ref{cc_R0}-\ref{cc_R3}) to determine the explicit moments for the nonequilibrium and inhomogeneous Holstein-Hubbard model. To simplify the formulas, we introduce the notation, $\Bar{O}={O}(T_{ave})$, to indicate the operator is evaluated at the average time $T_{ave}$, after taking the limit $t_{rel}\rightarrow0$. The overbar is also used to indicate a simple function of time is evaluated at $T_{ave}$. In addition, we define $\bar{\nu}_{i\sigma}=\mu_{i}(T_{ave})-U_{i}(T_{ave})\langle n_{i\bar\sigma}(T_{ave})\rangle-g_{i}(T_{ave})\langle x_{i}(T_{ave})\rangle$ to make the expressions more readable (we also use the notation $\bar\sigma=-\sigma$); be careful not to conflate the meaning of the overbar on the $\sigma$ label with the meaning of the overbar on an operator or a time-dependent function.
 
 After some significant algebra, we find the following results. The zeroth moment is trivial,
\begin{eqnarray}\label{HH_R0}\
\mu_{ij\sigma}^{R0}(T_{ave})=\delta_{ij},
\end{eqnarray}
and higher moments are shown below, where we employed the fermionic operator identity $n_{i\sigma}^2=n_{i\sigma}$ to simplify the final results
%
%
\begin{equation}\label{HH_R1}
\mu_{ij\sigma}^{R1}(T)=-\bar{t}_{ij}-\bar{\nu}_{i\sigma}\delta_{ij}
\end{equation}
%
%
\begin{eqnarray}\label{HH_R2}\
\mu_{ij\sigma}^{R2}(T_{ave})&=&\sum_{k}\bar{t}_{ik}\bar{t}_{kj}+\bar{t}_{ij}\bar{\nu}_{i\sigma}+
\bar{t}_{ij}\bar{\nu}_{j\sigma}+\bar{\nu}_{i\sigma}^{2}\delta_{ij}\nonumber\\
&+&
\bar{U}_{i}^{2}[\langle \bar{n}_{i\bar\sigma}\rangle-\langle \bar{n}_{i\bar\sigma}\rangle^{2}]\delta_{ij}+\bar{g}_{i}^{2}[\langle \bar{x}_{i}^{2}\rangle
-\langle \bar{x}_{i}\rangle^{2}]\delta_{ij}\nonumber\\
&+&2\bar{U}_{i}\bar{g}_{i}[\langle \bar{n}_{i\bar\sigma}\bar{x}_{i}\rangle-\langle \bar{n}_{i\bar\sigma}\rangle\langle \bar{x}_{i}\rangle]\delta_{ij},
\end{eqnarray}
The third moment is obviously of considerably greater complexity than the previous two. After a number of simplifications, we find 
\begin{widetext}
\begin{eqnarray}
    \label{HH_R3}
    \mu_{ij\sigma}^{R3}(T_{ave})= &&-{\rm Re}\left(\sum_{l s}\bar{t}_{il}\bar{t}_{l s}\bar{t}_{sj}\right) -{\rm Re}\left(\sum_{k}\bar{t}_{ik}\bar{t}_{kj}\left(\bar{\nu}_{i\sigma}+\bar{\nu}_{j\sigma}+\bar{\nu}_{k\sigma}\right)\right)-\delta_{ij}\bar{\nu}_{i\sigma}^{3}+\delta_{ij}\bar{g}_{i}^{3}\left[\left<\bar{x}_{i}^{3}\right>-\left<\bar{x}_{i}\right>^{3}\right]\nonumber \\
    &&-3\delta_{ij}\bar{\mu}_{i}\bar{g}_{i}^{2}\left[\left<\bar{x}_{i}^{2}\right>-\left<\bar{x}_{i}\right>^{2}\right]+\delta_{ij}\bar{U}_{i}^{3}\left[\left< \bar{n}_{i\bar{\sigma}}\right>-\left< \bar{n}_{i\bar{\sigma}}\right>^{3}\right]+3\delta_{ij}\bar{g}_{i}^{2}\bar{U}_{i}\left[\left<\bar{n}_{i\bar{\sigma}}\bar{x}^2\right>-\left<\bar{n}_{i\bar{\sigma}}\right>\left<\bar{x}_i\right>^2\right]\nonumber\\
    &&-6\delta_{ij}\bar{\mu}_{i}\bar{g}_{i}\bar{U}_{i}\left[\left<\bar{n}_{i\bar{\sigma}}\bar{x}_{i}\right>-\left<\bar{n}_{i\bar{\sigma}}\right>\left<\bar{x}_{i}\right>\right]+3\delta_{ij}\bar{g}_{i}\bar{U}_{i}^{2}\left[\left<\bar{n}_{i\bar{\sigma}}\bar{x}_{i}\right>-\left<\bar{n}_{i\bar{\sigma}}\right>^{2}\left<\bar{x}_{i}\right>\right]-3\delta_{ij}\bar{\mu}_{i}\bar{U}_{i}^{2}\left[\left<\bar{n}_{i\bar{\sigma}}\right>-\left<\bar{n}_{i\bar{\sigma}}\right>^{2}\right]\nonumber\\
    &&-{\rm Re}\left(t_{ij}\left(\bar{\nu}_{i\sigma}^{2}+\bar{g}_{i}^{2}\left[\left<\bar{x}_{i}^{2}\right>-\left<\bar{x}_{i}\right>^{2}\right]+2\bar{g}_{i}\bar{U}_{i}\left[\left<\bar{x}_{i}\bar{n}_{i\bar{\sigma}}\right>-\left<\bar{x}_{i}\right>\left<\bar{n}_{i\bar{\sigma}}\right>\right]+\bar{U}_{i}^{2}\left[\left<\bar{n}_{i\bar{\sigma}}\right>-\left<\bar{n}_{i\bar{\sigma}}\right>^{2}\right]\right)\right)\nonumber\\
    &&-{\rm Re}\left(t_{ij}\left(\bar{\nu}_{j\sigma}^{2}+\bar{g}_{j}^{2}\left[\left<\bar{x}_{j}^{2}\right>-\left<\bar{x}_{j}\right>^{2}\right]+2\bar{g}_{j}\bar{U}_{j}\left[\left<\bar{x}_{j}\bar{n}_{j\bar{\sigma}}\right>-\left<\bar{x}_{j}\right>\left<\bar{n}_{j\bar{\sigma}}\right>\right]+\bar{U}_{j}^{2}\left[\left<\bar{n}_{j\bar{\sigma}}\right>-\left<\bar{n}_{j\bar{\sigma}}\right>^{2}\right]\right)\right)\nonumber\\
    &&-{\rm Re}\left(t_{ij}\left(\bar{\nu}_{i\sigma}\bar{\nu}_{j\sigma}+\bar{g}_{i}\bar{g}_{j}\left[\left<\bar{x}_{i}\bar{x}_{j}\right>-\left<\bar{x}_{i}\right>\left<\bar{x}_{j}\right>\right]+\bar{U}_{i}\bar{U}_{j}\left[\left<\bar{n}_{i\bar{\sigma}}\bar{n}_{j\bar{\sigma}}\right>-\left<\bar{n}_{i\bar{\sigma}}\right>\left<\bar{n}_{j\bar{\sigma}}\right>\right]\right)\right)\nonumber\\
    &&-{\rm Re}\left(t_{ij}\left(\bar{g}_{i}\bar{U}_{j}\left[\left<\bar{x}_{i}\bar{n}_{j\bar{\sigma}}\right>-\left<\bar{x}_{i}\right>\left<\bar{n}_{j\bar{\sigma}}\right>\right]+\bar{g}_{j}\bar{U}_{i}\left[\left<\bar{x}_{j}\bar{n}_{i\bar{\sigma}}\right>-\left<\bar{x}_{j}\right>\left<\bar{n}_{i\bar{\sigma}}\right>\right]\right)\right)\nonumber \\
    &&+\delta_{ij}\frac{\bar{g}_{i}^{2}}{2m}\left[1-\frac{1}{2}\left(\left<\bar{n}_{i\sigma}\right>+\left<\bar{n}_{i\bar{\sigma}}\right>\right)\right]+\frac{1}{4} \delta_{ij}\frac{\kappa_{i}\bar{g}_{i}}{m_{i}} \left<x_{i}\right>+{\rm Re}\left(i  \bar{t}_{ij}\sum_{l}{\rm Im}\, \left(\bar{U}_{i}\bar{t}_{il}\left<\bar{c}^{\dagger}_{i \bar{\sigma}}\bar{c}_{l\bar{\sigma}}\right>-\bar{U}_{j}\bar{t}_{jl}\left<\bar{c}^{\dagger}_{j\bar{\sigma}}\bar{c}_{l\bar{\sigma}}\right>\right)\right) \nonumber \\
    &&-\frac{1}{2}\bar{U}_{i}\delta_{ij}\sum_{l s}{\rm Re}\, \left(\bar{t}_{l i}\bar{t}_{is}\left<\bar{c}_{l\bar{\sigma}}^{\dagger}\bar{c}_{s\bar{\sigma}}\right>-\bar{t}_{l i}\bar{t}_{sl}\left<\bar{c}_{s\bar{\sigma}}^{\dagger}\bar{c}_{i\bar{\sigma}}\right>\right)\nonumber \\
    &&+\frac{1}{2}\delta_{ij}\bar{U}_i\sum_{l}\left(\bar{g}_i\left< \bar{x}_{i}{\rm Re}\left(\bar{t}_{l i}\bar{c}^{\dagger}_{l\bar{\sigma}}\bar{c}_{i\bar{\sigma}}\right)\right> -\bar{g}_{l}\left< \bar{x}_{l}{\rm Re}\left(\bar{t}_{l i}\bar{c}^{\dagger}_{l\bar{\sigma}}\bar{c}_{i\bar{\sigma}}\right)\right>\right)\nonumber\\
    &&-\frac{1}{2}\bar{U}_i\delta_{ij}\sum_{l}\left(\bar{\mu}_{i}-\bar{\mu}_{l}\right){\rm Re}\left(\bar{t}_{l i}\left< \bar{c}^{\dagger}_{l\bar{\sigma}}\bar{c}_{i\bar{\sigma}}\right>\right)-\frac{1}{2}\bar{U}_i\delta_{ij}\sum_{l}\bar{U}_{l}\left< \bar{n}_{l\sigma}{\rm Re}\left(\bar{t}_{l i}\bar{c}^{\dagger}_{l\bar{\sigma}}\bar{c}_{i\bar{\sigma}}\right)\right>\nonumber \\
    &&+2\bar{U}_{i}\bar{U}_{j} \left<{\rm Re}\, (\bar{t}_{ij}\bar{c}_{i\bar{\sigma}}^{\dagger}\bar{c}_{j\bar{\sigma}})\bar{c}_{j\sigma}^{\dagger}\bar{c}_{i\sigma}\right>-\frac{3}{2}\delta_{ij}\bar{U}_{i}^{2}\sum_{l}\left<\bar{n}_{i\sigma}{\rm Re}\, \left(\bar{t}_{l i}\bar{c}^{\dagger}_{l\bar{\sigma}}\bar{c}_{i\bar{\sigma}}\right)\right> \nonumber \\
    &&+\frac{1}{2}{\rm Re}\,\left( i\sum_{k}\left[\frac{d\bar{t}_{ik}}{dT_{ave}}\bar{t}_{kj}-\bar{t}_{ik}\frac{d\bar{t}_{kj}}{dT_{ave}}\right]\right)-\frac{1}{2}{\rm Re}\,\left( i\frac{d\bar{t}_{ij}}{dT_{ave}}\left[\bar{\mu}_{i}-\bar{\mu}_{j}\right]\right)+\frac{1}{2}{\rm Re}\,\left(i\bar{t}_{ij}\left[\frac{d\bar{\mu}_{i}}{dT_{ave}}-\frac{d\bar{\mu}_{j}}{dT_{ave}}\right]\right)\nonumber\\
    &&+\frac{1}{2}{\rm Re}\, \left(i\frac{d\bar{t}_{ij}}{dT_{ave}}\left[\bar{g}_{i}\left< \bar{x}_{i}\right>-\bar{g}_{j}\left< \bar{x}_{j}\right>\right]\right)-\frac{1}{2}{\rm Re}\,\left(i\bar{t}_{ij}\left[\frac{d\bar{g}_{i}}{dT_{ave}}\left< \bar{x}_{i}\right>-\frac{d\bar{g}_{j}}{dT_{ave}}\left< \bar{x}_{j}\right>\right]\right)\nonumber\\
    &&+\frac{1}{2}{\rm Re}\, \left(i\frac{d\bar{t}_{ij}}{dT_{ave}}[\bar{U}_{i}\left< \bar{n}_{i\bar\sigma}\right>-\bar{U}_{j}\left< \bar{n}_{j\bar\sigma}\right>]\right)-\frac{1}{2}{\rm Re}\, \left(i\bar{t}_{ij}\left[\frac{d\bar{U}_{i}}{dT_{ave}}\left< \bar{n}_{i\bar\sigma}\right>-\frac{d\bar{U}_{j}}{dT_{ave}}\left< \bar{n}_{j\bar\sigma}\right>\right]\right)\nonumber\\
    &&-\frac{1}{2}\delta_{ij}\frac{d\bar{g}_i}{dT_{ave}}\frac{d\left<\bar{x}_{i}\right>}{dT_{ave}}+\delta_{ij}\frac{d\bar{U}_i}{dT_{ave}}\sum_{k}{\rm Im}\, \left(\bar{t}_{ik}\left< \bar{c}_{i\bar{\sigma}}^{\dagger}\bar{c}_{k\bar{\sigma}}\right>\right)-\delta_{ij}\bar{U}_{i}\sum_{k}{\rm Im}\left(\frac{d\bar{t}_{ik}}{dT_{ave}}\left< \bar{c}_{i\bar{\sigma}}^{\dagger}\bar{c}_{k\bar{\sigma}}\right>\right)\nonumber\\
    &&+\frac{1}{4}{\rm Re}\,\left(\frac{d^{2}\bar{t}_{ij}}{dT_{ave}^{2}}\right)+\frac{1}{4}\delta_{ij}{\rm Re}\,\left(\frac{d^2\bar{\mu}_{i}}{dT_{ave}^{2}}-\frac{d^2\bar{U}_{i}}{dT_{ave}^{2}}\left< \bar{n}_{i\bar\sigma}\right>-\frac{d^2\bar{g}_{i}}{dT_{ave}^{2}}\left< \bar{x}_{i}\right>\right)
\end{eqnarray}

%
\end{widetext}
These are the main results of this work. 

\section{Formalism for the sum rules for the retarded electronic self-energy}
 Next, we derive the retarded self-energy moments. The self-energy does not vanish at high frequency, but approaches a constant value, which we denote $\Sigma_{ij\sigma}^R(T_{ave},\omega=\infty)$ and which is real. The moments are defined from integrals over the imaginary part of the self-energy via
\begin{equation}
C_{ij\sigma}^{Rn}=-\frac{1}{\pi}\int d\omega \omega^n {\rm Im}\Sigma_{ij\sigma}(\omega).
\label{eq: se_moment_def}
\end{equation}
The zeroth moment gives the overall strength of the self-energy. These moments can be obtained from the Dyson equation, which connects the self-energy with the Green's function. For the nonequilibrium case, it is useful to work in the Larkin-Ovchinkov representation where the Green's function and the self-energy each become $2\times2$ matrices \cite{Larkin}. The complete derivation of this Dyson equation for the nonequilibrium self-energy is presented in Ref.~\onlinecite{Freericks2}, which we will quote and subsequently rearrange to extract the moments of the self-energy. Throughout this section we will employ a notation in which $\tilde{\mu}_{ij\sigma}^{Rn} =\left. \mu_{ij\sigma}^{Rn}\right|_{U=g=0}$. Otherwise stated, the tilde will denote the ``non-interacting'' case in which both $U_{i}(t)$ and $g_{i}(t)$ are zero. The identities are
\begin{eqnarray}\label{c_R0}\
\mu_{ij\sigma}^{R0}(T_{ave})=\tilde{\mu}_{ij\sigma}^{R0}(T_{ave}),
\end{eqnarray}
\begin{eqnarray}\label{c_R1}\
&~&\mu_{ij\sigma}^{R1}(T_{ave})=\tilde{\mu}_{ij\sigma}^{R1}(T_{ave})\nonumber\\
&+&\sum_{kl}\tilde{\mu}_{ik\sigma}^{R0}(T_{ave})\Sigma_{kl\sigma}^{R}(T_{ave},\omega=\infty)\mu_{lj\sigma}^{R0}(T_{ave}),
\end{eqnarray}
\begin{eqnarray}\label{c_R2}\
&~&\mu_{ij\sigma}^{R2}(T_{ave})=\tilde{\mu}_{ij\sigma}^{R2}(T_{ave})\nonumber\\
&+&\sum_{kl}\tilde{\mu}_{ik\sigma}^{R0}(T_{ave})\Sigma_{kl\sigma}^{R}(T_{ave},\omega=\infty)\mu_{lj\sigma}^{R1}(T_{ave})\nonumber \\
&+&\sum_{kl}\tilde{\mu}_{ik\sigma}^{R0}(T_{ave})C_{kl\sigma}^{R0}(T_{ave})\mu_{lj\sigma}^{R0}(T_{ave})\nonumber \\ &+&\sum_{kl}\tilde{\mu}_{ik\sigma}^{R1}(T_{ave})\Sigma_{kl\sigma}^{R}(T_{ave},\omega=\infty)\mu_{lj\sigma}^{R0}(T_{ave}),
\end{eqnarray}
%
\begin{eqnarray}\label{c_R3}\
&&\mu_{ij\sigma}^{R3}(T_{ave})=\tilde{\mu}_{ij\sigma}^{R3}(T_{ave})\nonumber\\
&&+\sum_{kl}\tilde{\mu}_{ik\sigma}^{R0}(T_{ave})\Sigma_{kl\sigma}^{R}(T_{ave},\omega=\infty)\mu_{lj\sigma}^{R2}(T_{ave})\nonumber \\
&&+\sum_{kl}\tilde{\mu}_{ik\sigma}^{R0}(T_{ave})C_{kl\sigma}^{R0}(T_{ave})\mu_{lj\sigma}^{R1}(T_{ave})\nonumber\\
&&+\sum_{kl}\tilde{\mu}_{ik\sigma}^{R0}(T_{ave})C_{kl\sigma}^{R1}(T_{ave})\mu_{lj\sigma}^{R0}(T_{ave})\nonumber \\ 
&&+\sum_{kl}\tilde{\mu}_{ik\sigma}^{R1}(T_{ave})\Sigma_{kl\sigma}^{R}(T_{ave},\omega=\infty)\mu_{lj\sigma}^{R1}(T_{ave})\nonumber\\
&&+\sum_{kl}\tilde{\mu}_{ik\sigma}^{R1}(T_{ave})C_{kl\sigma}^{R0}(T_{ave})\mu_{lj\sigma}^{R0}(T_{ave})\nonumber \\ 
&&+\sum_{kl}\tilde{\mu}_{ik\sigma}^{R2}(T_{ave})\Sigma_{kl}^{R}(T_{ave},\omega=\infty)\mu_{lj\sigma}^{R0}(T_{ave}),
\end{eqnarray}
%
%
where $\Sigma_{ij}^R(\omega=\infty)$ is the high-frequency limit of the self-energy, the real constant term of the self-energy. Using the fact that
\begin{eqnarray}\label{cHH_Rconst}\
\mu_{ij\sigma}^{R0}(T_{ave})=\tilde{\mu}_{ij\sigma}^{R0}(T_{ave})=\delta_{ij},
\end{eqnarray}
the self-energy moment sum rules can be explicitly determined after some algebra. We find
\begin{equation}\label{CHH_R0}\
\Sigma^R_{ij\sigma}(T_{ave},\omega=\infty)=[\bar{U}_{i}\langle \bar{n}_{i\bar\sigma}\rangle +\bar{g}_{i}\langle \bar{x}_i\rangle] \delta_{ij},
\end{equation}
\begin{eqnarray}\label{CHH_R1}\
C^{R0}_{ij\sigma}(T_{ave})&=&\bar{U}_{i}^2[\langle \bar{n}_{i\bar\sigma}\rangle -\langle \bar{n}_{i\bar\sigma}\rangle^2]\delta_{ij}+\bar{g}_{i}^2[\langle \bar{x}_i^2\rangle -\langle \bar{x}_i\rangle^2]\delta_{ij}\nonumber\\
&+&2\bar{g}_{i}\bar{U}_{i}[\langle \bar{x}_{i}\bar{n}_{i\bar\sigma}\rangle-\langle \bar{x}_{i}\rangle\langle \bar{n}_{i\bar\sigma}\rangle]\delta_{ij}
\end{eqnarray}
and
\begin{widetext}
\begin{eqnarray}
    \label{CHH_R2}
    C^{R1}_{ij\sigma}(T_{ave})&=& \delta_{ij}\bar{U}_{i}^{3}\left[3\left<\bar{n}_{i\bar{\sigma}}\right>^{3}-2\left<\bar{n}_{i\bar{\sigma}}\right>^{2}+\left<\bar{n}_{i\bar{\sigma}}\right>\right]+\delta_{ij}\bar{g}_{i}^{3}\left[\left<\bar{x}_{i}^{3}\right>-2\left<\bar{x}_{i}^{2}\right>\left<\bar{x}_{i}\right>+\left<\bar{x}_{i}\right>^{3}\right] \nonumber \\
    && +\delta_{ij} \left[\bar{g}_{i}\bar{U}_{i}^{2}+\bar{g}_{i}^{2}\bar{U}_{i}\right] \left[3 \left<\bar{n}_{i\bar{\sigma}}\bar{x}_{i}^{2}\right>-2\left<\bar{n}_{i\bar{\sigma}}\right>\left<\bar{x}_{i}^{2}\right>-4\left<\bar{n}_{i\bar{\sigma}}\bar{x}_{i}\right>\left<\bar{x}_{i}\right>+3\left<\bar{n}_{i\bar{\sigma}}\right>\left<\bar{x}_{i}\right>^{2}\right]\nonumber \\
    && -2\delta_{ij}\bar{\mu}_{i}\bar{g}_{i}\bar{U}_{i}\left[\left<\bar{n}_{i\bar{\sigma}}\bar{x}_{i}\right>-\left<\bar{n}_{i\bar{\sigma}}\right>\left<\bar{x}_{i}\right>\right]-\delta_{ij}\bar{\mu}_{i}\bar{U}_{i}^{2}\left[\left<\bar{n}_{i\bar{\sigma}}\right>-\left<\bar{n}_{i\bar{\sigma}}\right>^{2}\right]-\delta_{ij}\bar{\mu}_{i}\bar{g}_{i}^{2}\left[\left<\bar{x}_{i}^{2}\right>-\left<\bar{x}_{i}\right>^{2}\right]\nonumber \\
    && -\bar{t}_{ij}\bar{g}_{i}\bar{g}_{j}\left[\left<\bar{x}_{i}\bar{x}_{j}\right>-\left<\bar{x}_{i}\right>\left<\bar{x}_{j}\right>\right]-\bar{t}_{ij}\bar{U}_{i}\bar{U}_{j}\left[\left<\bar{n}_{i\bar{\sigma}}\bar{n}_{j\bar{\sigma}}\right>-\left<\bar{n}_{i\bar{\sigma}}\right>\left<\bar{n}_{j\bar{\sigma}}\right>\right] \nonumber \\
    && -\bar{t}_{ij}\bar{g}_{i}\bar{U}_{j}\left[\left<\bar{n}_{j\bar{\sigma}}\bar{x}_{i}\right>-\left<\bar{n}_{j\bar{\sigma}}\right>\left<\bar{x}_{i}\right>\right]-\bar{t}_{ij}\bar{g}_{j}\bar{U}_{i}\left[\left<\bar{n}_{i\bar{\sigma}}\bar{x}_{j}\right>-\left<\bar{n}_{i\bar{\sigma}}\right>\left<\bar{x}_{j}\right>\right]-\bar{t}_{ij}\bar{\mu}_{i}^{2}-\bar{t}_{ij}\bar{\mu}_{j}^{2}\nonumber \\
    &&+\delta_{ij}\frac{\bar{g}_{i}^{2}}{2m}\left[1-\frac{1}{2}\left(\left<\bar{n}_{i\sigma}\right>+\left<\bar{n}_{i\bar{\sigma}}\right>\right)\right]+\frac{1}{4} \delta_{ij}\frac{\kappa_{i}\bar{g}_{i}}{m_{i}} \left<\bar{x}_{i}\right>+{\rm Re}\left(i  \bar{t}_{ij}\sum_{l}{\rm Im}\, \left(\bar{U}_{i}\bar{t}_{il}\left<\bar{c}^{\dagger}_{i \bar{\sigma}}\bar{c}_{l\bar{\sigma}}\right>-\bar{U}_{j}\bar{t}_{jl}\left<\bar{c}^{\dagger}_{j\bar{\sigma}}\bar{c}_{l\bar{\sigma}}\right>\right)\right) \nonumber \\
    &&-\frac{1}{2}\bar{U}_{i}\delta_{ij}\sum_{l s}{\rm Re}\, \left(\bar{t}_{l i}\bar{t}_{is}\left<\bar{c}_{l\bar{\sigma}}^{\dagger}\bar{c}_{s\bar{\sigma}}\right>-\bar{t}_{l i}\bar{t}_{sl}\left<\bar{c}_{s\bar{\sigma}}^{\dagger}\bar{c}_{i\bar{\sigma}}\right>\right)\nonumber \\
    &&+\frac{1}{2}\delta_{ij}\bar{U}_i\sum_{l}\left(\bar{g}_i\left< \bar{x}_{i}{\rm Re}\left(\bar{t}_{l i}\bar{c}^{\dagger}_{l\bar{\sigma}}\bar{c}_{i\bar{\sigma}}\right)\right> -\bar{g}_{l}\left< \bar{x}_{l}{\rm Re}\left(\bar{t}_{l i}\bar{c}^{\dagger}_{l\bar{\sigma}}\bar{c}_{i\bar{\sigma}}\right)\right>\right)\nonumber\\
    &&-\frac{1}{2}\bar{U}_i\delta_{ij}\sum_{l}\left(\bar{\mu}_{i}-\bar{\mu}_{l}\right){\rm Re}\left(\bar{t}_{l i}\left< \bar{c}^{\dagger}_{l\bar{\sigma}}\bar{c}_{i\bar{\sigma}}\right>\right)-\frac{1}{2}\bar{U}_i\delta_{ij}\sum_{l}\bar{U}_{l}\left< \bar{n}_{l\sigma}{\rm Re}\left(\bar{t}_{l i}\bar{c}^{\dagger}_{l\bar{\sigma}}\bar{c}_{i\bar{\sigma}}\right)\right>\nonumber \\
    &&+2\bar{U}_{i}\bar{U}_{j} \left<{\rm Re}\, (\bar{t}_{ij}\bar{c}_{i\bar{\sigma}}^{\dagger}\bar{c}_{j\bar{\sigma}})\bar{c}_{j\sigma}^{\dagger}\bar{c}_{i\sigma}\right>-\frac{3}{2}\delta_{ij}\bar{U}_{i}^{2}\sum_{l}\left<\bar{n}_{i\sigma}{\rm Re}\, \left(\bar{t}_{l i}\bar{c}^{\dagger}_{l\bar{\sigma}}\bar{c}_{i\bar{\sigma}}\right)\right> \nonumber \\
    &&+\frac{1}{2}{\rm Re}\, \left(i\frac{d\bar{t}_{ij}}{dT_{ave}}\left[\bar{g}_{i}\left< \bar{x}_{i}\right>-\bar{g}_{j}\left< \bar{x}_{j}\right>\right]\right)-\frac{1}{2}{\rm Re}\,\left(i\bar{t}_{ij}\left[\frac{d\bar{g}_{i}}{dT_{ave}}\left< \bar{x}_{i}\right>-\frac{d\bar{g}_{j}}{dT_{ave}}\left< \bar{x}_{j}\right>\right]\right)\nonumber\\
    &&+\frac{1}{2}{\rm Re}\, \left(i\frac{d\bar{t}_{ij}}{dT_{ave}}[\bar{U}_{i}\left< \bar{n}_{i\bar\sigma}\right>-\bar{U}_{j}\left< \bar{n}_{j\bar\sigma}\right>]\right)-\frac{1}{2}{\rm Re}\, \left(i\bar{t}_{ij}\left[\frac{d\bar{U}_{i}}{dT_{ave}}\left< \bar{n}_{i\bar\sigma}\right>-\frac{d\bar{U}_{j}}{dT_{ave}}\left< \bar{n}_{j\bar\sigma}\right>\right]\right)\nonumber\\
    &&-\frac{1}{2}\delta_{ij}\frac{d\bar{g}_i}{dT_{ave}}\frac{d\left<\bar{x}_{i}\right>}{dT_{ave}}+\delta_{ij}\frac{d\bar{U}_i}{dT_{ave}}\sum_{k}{\rm Im}\, \left(\bar{t}_{ik}\left< \bar{c}_{i\bar{\sigma}}^{\dagger}\bar{c}_{k\bar{\sigma}}\right>\right)-\delta_{ij}\bar{U}_{i}\sum_{k}{\rm Im}\left(\frac{d\bar{t}_{ik}}{dT_{ave}}\left< \bar{c}_{i\bar{\sigma}}^{\dagger}\bar{c}_{k\bar{\sigma}}\right>\right)\nonumber\\
    &&-\frac{1}{4}\delta_{ij}{\rm Re}\,\left(\frac{d\bar{U}_{i}^{2}}{dT_{ave}^{2}}\left< \bar{n}_{i\bar\sigma}\right>+\frac{d\bar{g}_{i}^{2}}{dT_{ave}^{2}}\left< \bar{x}_{i}\right>\right)
\end{eqnarray}
%

\end{widetext}

Note that the zeroth moment is local (diagonal) even if the self-energy has momentum dependence, while the first moment can be nonzero only for local terms ($i=j$) and for terms where the hopping is nonvanishing ($t_{ij}(T_{ave})\ne 0$). In particular, if we use the zeroth moment to determine the strength of the effective electron-phonon interaction, then for a pure Holstein model, the only way the electron-phonon interaction is dynamically changed is if the correlation function of the phonon coordinate changes as a function of time. This can happen, for example, if energy flows into the phonon bath, but is likely to be delayed due to the bottleneck for energy flow from electrons to phonons. Screening effects, which can also change the net electron-phonon coupling, are not in the Holstein-Hubbard model, and require a more complex model to be properly described.

\section{\label{sec:level2} Spectral sum rules in momentum space}
When the system is translationally invariant, it is convenient to work in momentum space. Thus, we examine the case where $t_{ij}$ is a periodic hopping matrix and the local chemical potential, electron-phonon coupling, and Hubbard interaction are all spatially uniform. This calculation requires us to make an appropriate Fourier transformation. We start with the definition of the retarded Green's function in momentum space,
\begin{eqnarray}\label{GR_k1}\
G_{\bf k\sigma}^{R}(t,t')=-i\theta(t{-}t')\langle\{c_{\bf k\sigma}(t),c_{\bf k\sigma}(t')\}\rangle,
\end{eqnarray}
where ${\bf k}$ denotes the momentum.
The corresponding creation and annihilation operators in momentum space can be obtained by performing a Fourier transform, $c_{\bf k\sigma}=\sum_{i}e^{i\textbf{k}\cdot\textbf{R}_i}c_{i\sigma}/N$, and $c_{\bf k\sigma}^{\dagger}=
\sum_{i}e^{-i\textbf{k}\cdot\textbf{R}_i}c_{i\sigma}^{\dagger}/N$. Here, $N$ is the number of lattice sites. Substituting the inverse Fourier transformation into the formula for the real-space moments yields
\begin{eqnarray}\label{Lnm3}\
\mu_{\textbf{k}\sigma}^{Rn}(T_{ave})=\frac{1}{N}\sum_{ij}e^{-i\textbf{k}\cdot (\textbf{R}_{i}-\textbf{R}_{j})}\mu_{ij\sigma}^{Rn}(T_{ave}).
\end{eqnarray}
Now, all that is left is a tedious calculation. One simplification is particularly important, however: due to translational invariance, all strictly local expectation values are independent of the lattice site and can be replaced by site-independent numbers. Thus, terms like $\left<\bar{x}_{i}\bar{n}_{i\bar{\sigma}}\right>$ will be denoted $\left<\bar{x}\bar{n}_{\bar{\sigma}}\right>$, without spatial indices. At first this may seem a bit confusing, since local expectation values, such as the double occupancy, are typically written in terms of a sum over three independent momenta of a expectation value involving four fermion operators at different momenta. Rather than expressing the expectation values in this form, when they are constants, independent of the lattice site, we keep them in the local representation in the formulas summarized below. This greatly simplifies both the formulas and using them to numerically determine the moments.

While it is true that spatial indices can be suppressed for local expectation values without a loss of generality, this is not true for non-local expectation values because these terms will have explicit momentum dependence after Fourier transformation.

The momentum-based sum rules then become
\begin{eqnarray}\label{mu_k0}\
\mu_{\textbf{k}\sigma}^{R0}(T_{ave})=1,
\end{eqnarray}
\begin{eqnarray}\label{mu_k01}\
\mu_{\textbf{k}\sigma}^{R1}(T_{ave})=\bar{\epsilon}_{\textbf{k}}-\bar{\nu}_\sigma,
\end{eqnarray}
 where $\bar{\epsilon}_\textbf{k}=-\sum_{\{\boldsymbol{\delta}\}}t_{i\,i+\delta}\left(T_{ave}\right)e^{i\textbf{k}\cdot\boldsymbol{\delta}}$,
$\{\boldsymbol{\delta}\}$
 
is the set of all of the translation vectors for which the hopping matrix is nonzero (the index $i+\delta$ schematically denotes the lattice site corresponding to site $\textbf{ R}_i+\boldsymbol{\delta}$), and $\bar{\nu}_\sigma=\mu(T_{ave})-U(T_{ave})\langle n_{\bar\sigma}(T_{ave})\rangle-g(T_{ave})\langle x(T_{ave})\rangle$. Note, that in a paramagnetic solution, the filling will be independent of the spin $\sigma$. We also must define the momentum space phonon position operator, $x_{\mathbf{k}}$, which is defined as simply the Fourier transform of the position space operator: $x_{\mathbf{k}} = \frac{1}{N}\sum_{i}e^{-i \mathbf{k}\cdot \mathbf{R}_{i}}x_{i}$. The higher moments become the following:

\begin{widetext}
\begin{equation}\label{mu_k02}\
\mu_{\textbf{k}\sigma}^{R2}(T_{ave})=\bar{\epsilon}_{\textbf{k}}^{2}-2\bar{\epsilon}_{\textbf{k}}\bar{\nu}_\sigma+\bar{\nu}_\sigma^{2}+\bar{U}^{2}[\langle \bar{n}_{\bar\sigma}\rangle-\langle \bar{n}_{\bar\sigma}\rangle^{2}]
+\bar{g}^{2}[\langle \bar{x}^{2}\rangle-\langle \bar{x}\rangle^{2}]+2\bar{U}\bar{g}[\langle \bar{n}_{\bar\sigma}\bar{x}\rangle-\langle \bar{n}_{\bar\sigma}\rangle\langle \bar{x}\rangle],
\end{equation}
\begin{eqnarray}
\label{mu_k03}
    \mu_{\mathbf{k}\sigma}^{R3}(T_{ave})= &&\bar{\epsilon}_{\mathbf{k}}^{3}- 3\bar{\epsilon}_{\mathbf{k}}^{2}\bar{\nu}_{\sigma}-\bar{\nu}_{\sigma}^{3}+\bar{g}^{3}\left[\left<\bar{x}^{3}\right>-\left<\bar{x}\right>^{3}\right]\nonumber \\
    &&-3\bar{\mu}\bar{g}^{2}\left[\left<\bar{x}^{2}\right>-\left<\bar{x}\right>^{2}\right]+\bar{U}^{3}\left[\left< \bar{n}_{\bar{\sigma}}\right>-\left< \bar{n}_{\bar{\sigma}}\right>^{3}\right]+3\bar{g}^{2}\bar{U}\left[\left<\bar{n}_{\bar{\sigma}}\bar{x}^2\right>-\left<\bar{n}_{\bar{\sigma}}\right>\left<\bar{x}\right>^2\right]\nonumber\\
    &&-6\bar{\mu}\bar{g}\bar{U}\left[\left<\bar{n}_{\bar{\sigma}}\bar{x}\right>-\left<\bar{n}_{\bar{\sigma}}\right>\left<\bar{x}\right>\right]+3\bar{g}\bar{U}^{2}\left[\left<\bar{n}_{\bar{\sigma}}\bar{x}\right>-\left<\bar{n}_{\bar{\sigma}}\right>^{2}\left<\bar{x}\right>\right]-3\bar{\mu}\bar{U}^{2}\left[\left<\bar{n}_{\bar{\sigma}}\right>-\left<\bar{n}_{\bar{\sigma}}\right>^{2}\right]\nonumber\\
    &&-2\bar{\epsilon}_{\mathbf{k}}\left(\bar{g}^{2}\left[\left<\bar{x}^{2}\right>-\left<\bar{x}\right>^{2}\right]+2\bar{g}\bar{U}\left[\left<\bar{x}\bar{n}_{\bar{\sigma}}\right>-\left<\bar{x}\right>\left<\bar{n}_{\bar{\sigma}}\right>\right]+\bar{U}^{2}\left[\left<\bar{n}_{\bar{\sigma}}\right>-\left<\bar{n}_{\bar{\sigma}}\right>^{2}\right]\right)+3\bar{\epsilon}_{\mathbf{k}}\bar{\nu}_{\sigma}^{2}\nonumber\\
    &&-{\rm Re}\left(\bar{g}^{2}\left[\sum_{\mathbf{p}}\bar{\epsilon}_{\mathbf{k}+\mathbf{p}}\left<\bar{x}_{\mathbf{p}}\bar{x}_{-\mathbf{p}}\right>-\bar{\epsilon}_{\mathbf{k}}\left<\bar{x}\right>^{2}\right]+\bar{U}^{2}\left[\sum_{\mathbf{p}_{1-3}}\bar{\epsilon}_{\mathbf{p}_{1}}\left<\bar{c}^{\dagger  }_{\mathbf{p}_{2}\bar{\sigma}} \bar{c}_{\mathbf{p}_{1}+\mathbf{p}_{2}-\mathbf{k}\, \bar{\sigma}} \bar{c}^{\dagger }_{\mathbf{p}_{3}\bar{\sigma}}\bar{c}_{\mathbf{k}+\mathbf{p}_{3}-\mathbf{p}_{1} \bar{\sigma}}\right>-\bar{\epsilon}_{\mathbf{k}}\left<\bar{n}_{\bar{\sigma}}\right>^{2}\right]\right)\nonumber\\
    && -{\rm Re}\left(\bar{g}\bar{U}\left[\sum_{\mathbf{p}_{1-2}} \left(\bar{\epsilon}_{\mathbf{k}+\mathbf{p}_{1}}+\bar{\epsilon}_{\mathbf{k}-\mathbf{p}_{1}}\right)\left<x_{\mathbf{p}_{1}}\bar{c}^{\dagger}_{\mathbf{p}_{2}\bar{\sigma}}\bar{c}_{\mathbf{p}_{2}-\mathbf{p}_{1}\bar{\sigma}}\right>-2\bar{\epsilon}_{\mathbf{k}}\left<x\right>\left<n_{\bar{\sigma}}\right>\right]\right) \nonumber \\
    &&-\frac{1}{2}\bar{U}^{2}\sum_{\mathbf{p}_{1-3}}\bar{\epsilon}_{\mathbf{p}_{2}-\mathbf{p}_{3}-\mathbf{p}_{1}}\left< \bar{c}^{\dagger}_{\mathbf{p}_{1}\sigma}\bar{c}_{\mathbf{p}_{2}\sigma}{\rm Re}\left(\bar{c}^{\dagger}_{\mathbf{p}_{3}\bar{\sigma}}\bar{c}_{\mathbf{p}_{3}+\mathbf{p}_{1}-\mathbf{p}_{2}\bar{\sigma}}\right)\right>-\frac{3}{2}\bar{U}^{2}\sum_{\mathbf{p}_{1-3}}\bar{\epsilon}_{\mathbf{p}_{3}}\left< \bar{c}^{\dagger}_{\mathbf{p}_{1}\sigma}\bar{c}_{\mathbf{p}_{2}\sigma}{\rm Re}\left(\bar{c}^{\dagger}_{\mathbf{p}_{2}+\mathbf{p}_{3}-\mathbf{p}_{1}\bar{\sigma}}\bar{c}_{\mathbf{p}_{3}\bar{\sigma}}\right)\right> \nonumber \\
    && +2\bar{U}^{2} \sum_{\mathbf{p}_{1-3}}\bar{\epsilon}_{\mathbf{p}_{1}}\left<{\rm Re}\, \left(\bar{c}^{\dagger}_{\mathbf{p}_{2}\bar{\sigma}}\bar{c}_{\mathbf{p}_{3}\bar{\sigma}}\right))\bar{c}_{ \mathbf{p}_{3}-\mathbf{p}_{1}\sigma}^{\dagger}\bar{c}_{\mathbf{p}_{2}-\mathbf{p}_{1}\sigma}\right> +\frac{\bar{g}^{2}}{2m}\left[1-\frac{1}{2}\left(\left<\bar{n}_{\sigma}\right>+\left<\bar{n}_{\bar{\sigma}}\right>\right)\right]+\frac{\kappa\bar{g}}{4m} \left<\bar{x}\right>\nonumber \\
    &&-\frac{1}{2}\frac{d\bar{g}}{dT_{ave}}\frac{d\left<\bar{x}\right>}{dT_{ave}}+\frac{1}{4}{\rm Re}\,\left(\frac{d^{2}\bar{\epsilon}_{\mathbf{k}}}{dT_{ave}^{2}}\right)+\frac{1}{4}{\rm Re}\,\left(\frac{d\bar{\mu}^{2}}{dT_{ave}^{2}}-\frac{d\bar{U}^{2}}{dT_{ave}^{2}}\left< \bar{n}_{\bar\sigma}\right>-\frac{d\bar{g}^{2}}{dT_{ave}^{2}}\left< \bar{x}\right>\right).
\end{eqnarray}
\end{widetext}
Similarly, we can obtain the sum rules for the retarded self-energy in momentum space,
\begin{equation}\label{C_kconst}\
\Sigma^R_{\textbf{k}\sigma}(T_{ave},\omega=\infty)=\bar{U}\langle \bar{n}_{\bar\sigma}\rangle+\bar{g}\langle \bar{x}\rangle,
\end{equation}
%
%
\begin{eqnarray}\label{C_k0}\
C^{R0}_{\textbf{k}\sigma}(T_{ave})&=&\bar{U}^2[\langle \bar{n}_{\bar{\sigma}}\rangle -\langle \bar{n}_{\bar{\sigma}}\rangle^2]+\bar{g}^2[\langle \bar{x}^2\rangle-\langle \bar{x}\rangle^2]\nonumber \\
&& +2\bar{g}\bar{U}[\langle \bar{x}\bar{n}_{\bar{\sigma}}\rangle-\langle 
\bar{x}\rangle\langle \bar{n}_{\bar{\sigma}}\rangle]
\end{eqnarray}
\begin{widetext}
\begin{eqnarray}
\label{C_k1}\
    C^{R1}_{\mathbf{k}\sigma}(T_{ave})&=& \bar{U}^{3}\left[3\left<\bar{n}_{\bar{\sigma}}\right>^{3}-2\left<\bar{n}_{\bar{\sigma}}\right>^{2}+\left<\bar{n}_{\bar{\sigma}}\right>\right]+\bar{g}^{3}\left[\left<\bar{x}^{3}\right>-2\left<\bar{x}^{2}\right>\left<\bar{x}\right>+\left<\bar{x}\right>^{3}\right] \nonumber \\
    && + \left[\bar{g}\bar{U}^{2}+\bar{g}^{2}\bar{U}\right] \left[3 \left<\bar{n}_{\bar{\sigma}}\bar{x}_{}^{2}\right>-2\left<\bar{n}_{\bar{\sigma}}\right>\left<\bar{x}^{2}\right>-4\left<\bar{n}_{\bar{\sigma}}\bar{x}\right>\left<\bar{x}\right>+3\left<\bar{n}_{\bar{\sigma}}\right>\left<\bar{x}\right>^{2}\right]\nonumber \\
    && -2\bar{\mu}\bar{g}\bar{U}\left[\left<\bar{n}_{\bar{\sigma}}\bar{x}\right>-\left<\bar{n}_{\bar{\sigma}}\right>\left<\bar{x}\right>\right]-\bar{\mu}\bar{U}^{2}\left[\left<\bar{n}_{\bar{\sigma}}\right>-\left<\bar{n}_{\bar{\sigma}}\right>^{2}\right]-\bar{\mu}\bar{g}^{2}\left[\left<\bar{x}^{2}\right>-\left<\bar{x}\right>^{2}\right]\nonumber \\
    &&-{\rm Re}\left(\bar{g}^{2}\left[\sum_{\mathbf{p}}\bar{\epsilon}_{\mathbf{k}+\mathbf{p}}\left<\bar{x}_{\mathbf{p}}\bar{x}_{-\mathbf{p}}\right>-\bar{\epsilon}_{\mathbf{k}}\left<\bar{x}\right>^{2}\right]+\bar{U}^{2}\left[\sum_{\mathbf{p}_{1-3}}\bar{\epsilon}_{\mathbf{p}_{1}}\left<\bar{c}^{\dagger  }_{\mathbf{p}_{2}\bar{\sigma}} \bar{c}_{\mathbf{p}_{1}+\mathbf{p}_{2}-\mathbf{k}\, \bar{\sigma}} \bar{c}^{\dagger }_{\mathbf{p}_{3}\bar{\sigma}}\bar{c}_{\mathbf{k}+\mathbf{p}_{3}-\mathbf{p}_{1} \bar{\sigma}}\right>-\bar{\epsilon}_{\mathbf{k}}\left<\bar{n}_{\bar{\sigma}}\right>^{2}\right]\right) \nonumber \\
    && -{\rm Re}\left(\bar{g}\bar{U}\left[\sum_{\mathbf{p}_{1-2}} \left(\bar{\epsilon}_{\mathbf{k}+\mathbf{p}_{1}}+\bar{\epsilon}_{\mathbf{k}-\mathbf{p}_{1}}\right)\left<x_{\mathbf{p}_{1}}\bar{c}^{\dagger}_{\mathbf{p}_{2}\bar{\sigma}}\bar{c}_{\mathbf{p}_{2}-\mathbf{p}_{1}\bar{\sigma}}\right>-2\bar{\epsilon}_{\mathbf{k}}\left<x\right>\left<n_{\bar{\sigma}}\right>\right]\right)-2\bar{\mu}^{2}\bar{\epsilon}_{\mathbf{k}} \nonumber \\
    &&-\frac{1}{2}\bar{U}^{2}\sum_{\mathbf{p}_{1-3}}\bar{\epsilon}_{\mathbf{p}_{2}-\mathbf{p}_{3}-\mathbf{p}_{1}}\left< \bar{c}^{\dagger}_{\mathbf{p}_{1}\sigma}\bar{c}_{\mathbf{p}_{2}\sigma}{\rm Re}\left(\bar{c}^{\dagger}_{\mathbf{p}_{3}\bar{\sigma}}\bar{c}_{\mathbf{p}_{3}+\mathbf{p}_{1}-\mathbf{p}_{2}\bar{\sigma}}\right)\right>-\frac{3}{2}\bar{U}^{2}\sum_{\mathbf{p}_{1-3}}\bar{\epsilon}_{\mathbf{p}_{3}}\left< \bar{c}^{\dagger}_{\mathbf{p}_{1}\sigma}\bar{c}_{\mathbf{p}_{2}\sigma}{\rm Re}\left(\bar{c}^{\dagger}_{\mathbf{p}_{2}+\mathbf{p}_{3}-\mathbf{p}_{1}\bar{\sigma}}\bar{c}_{\mathbf{p}_{3}\bar{\sigma}}\right)\right> \nonumber \\
    && +2\bar{U}^{2} \sum_{\mathbf{p}_{1-3}}\bar{\epsilon}_{\mathbf{p}_{1}}\left<{\rm Re}\, \left(\bar{c}^{\dagger}_{\mathbf{p}_{2}\bar{\sigma}}\bar{c}_{\mathbf{p}_{3}\bar{\sigma}}\right))\bar{c}_{ \mathbf{p}_{3}-\mathbf{p}_{1}\sigma}^{\dagger}\bar{c}_{\mathbf{p}_{2}-\mathbf{p}_{1}\sigma}\right> +\frac{\bar{g}^{2}}{2m}\left[1-\frac{1}{2}\left(\left<\bar{n}_{\sigma}\right>+\left<\bar{n}_{\bar{\sigma}}\right>\right)\right]+\frac{\kappa\bar{g}}{4m} \left<\bar{x}\right>\nonumber \\
    &&-\frac{1}{2}\frac{d\bar{g}}{dT_{ave}}\frac{d\left<\bar{x}\right>}{dT_{ave}}-\frac{1}{4}{\rm Re}\,\left(\frac{d\bar{U}^{2}}{dT_{ave}^{2}}\left< \bar{n}_{\bar\sigma}\right>+\frac{d\bar{g}^{2}}{dT_{ave}^{2}}\left< \bar{x}\right>\right)
\end{eqnarray}
\end{widetext}

These forms of the  sum rules are more useful for calculations that work with translationally invariant systems. Note that, as expected, the moments either have no momentum dependence, or inherit a momentum dependence from the bandstructure, because the off-diagonal moments always had a dependence on the hopping matrix element. As noted before, the higher moments require many different expectation values to be known in order to properly employ them. If one is using methods like quantum Monte Carlo simulation, where one can measure such expectation values in addition to determining the Green's function and self-energy, then one can employ these results as a check on the accuracy of the calculations. Similarly, if one has an approximation method that is employed for the Holstein-Hubbard model, then by calculating these different expectation values within the approximation, one can test the overall self-consistency of the approximation to see if it satisfies these exact relations. Of course, if everything is evaluated with an approximate solution, there is no guarantee that the approximation is accurate even if it self-consistently satisfies these sum rules. But if the result does not satisfy the sum rules, it can be immediately falsified.

We also want to emphasize that these results hold in a wide range of different nonequilibrium situations and are quite general. Since there are only a few exact results known about nonequilibrium solutions, we hope the community will regularly use these sum rules to check the accuracy of different calculations, especially those in nonequilibrium.

\section{Verification of the Moments}
We have not been able to find sufficiently accurate numerical calculations on the full (spatially inhomogeneous and time-dependent) Holstein-Hubbard model, along with the calculation of the required expectation values, to compare the results of these sum rules against state-of-the-art numerical calculations. Instead, in our first check, we examine the so-called atomic limit, which allows us to check the pieces of the sum rule that do not depend on the hopping. The atomic limit, defined by $t_{ij}\to 0$, of the Holstein-Hubbard model represents a highly non-trivial interacting non-equilibrium system which admits an exact solution.\cite{Nesselrodt}~
This makes it an ideal candidate to verify the moments presented in Eqs.~(\ref{HH_R0}-\ref{HH_R3}) when the hopping vanishes. 

\begin{figure}[h]
    \centering
    \includegraphics[width = \linewidth]{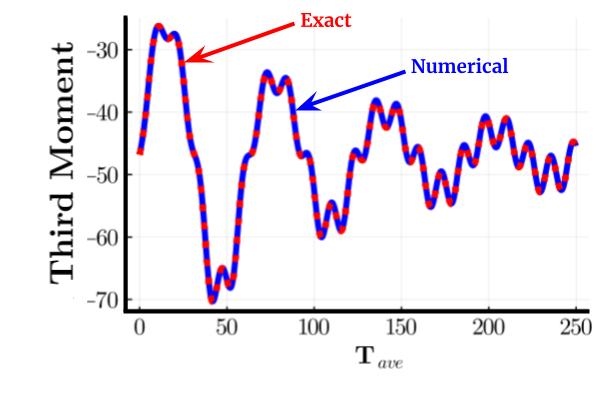}
    \caption{Numeric third derivative of the atomic limit Green's function compared to the exact results for the third moment (Eq.~(\ref{HH_R3}), $t_{rs}\to 0$), with time-dependent couplings\\ $g(t)=g_0+g_1\sin(at)e^{-bt}$ and $U(t)=U_0+U_1\sin(ct)$.}
    \label{fig:numericalvalidation}
\end{figure}

Using this exact solution of the non-equilibrium retarded Green's function for the model allows us to evaluate all the expectation values which appear in Eqs.~(\ref{HH_R0}-\ref{HH_R3}), when the hopping is set to zero. Then, there are two independent checks to verify the first four moment sum rules. First, uses a numerical differentiation of the Green's function, as prescribed in Eq.~(\ref{mu_t2}), and then plots these numerical derivative based moments against the exact calculation of the moments given by Eqs.~(\ref{HH_R0}-\ref{HH_R3}) determined by exactly evaluating the different expectation values. We do this as a function of $T_{ave}$, when both the electron-electron and electron-phonon couplings vary as a function of time, and find excellent agreement, which we show in Fig.~\ref{fig:numericalvalidation}. Second, by taking derivatives of the exact expressions for the Green's function by hand, again as described in Eq.~(\ref{mu_t2}), and comparing against the expressions for the expectation values which appear in the moments, an exact analytic verification of the first four spectral moments is found (for details see Ref.~\onlinecite{Nesselrodt}).

While not a complete verification of the spectral moments presented here (for example, there is no such check available for the moments of the self-energy or momentum space results), the evidence from the atomic limit of the model represents the best check of our results currently available, and in this case we find perfect agreement. 

It is, of course, also necessary to examine the extent of agreement with the literature. Consistency of the ${\rm zeroth}-{\rm second}$ order results is apparent throughout the relevant body of work\cite{Freericks1,Freericks1+,Freericks2,Freericks2+,Freericks3,Freericks3+,Freericks4,Freericks5}. For the $3{\rm rd}$ order results, verification is more complicated as Eqs.~(\ref{ccc_R2}-\ref{ccc_R3}) differ from previous results, through modifications of the coefficients of a few terms. This is because the older results have some errors, as we now elaborate. Note that the older results were checked against numerical calculations as well, but they could not be checked for all possible cases and, indeed, the errors appear in terms that could not be checked against numerical data.

We now discuss the issues with the previous work. Our results agree with the earlier position space results\cite{Freericks3,Freericks3+} with some modified coefficients on the ${\rm sixth}-{\rm eighth}$ lines and in the derivative terms of the result for the third moment as presented in the relevant erratum\cite{Freericks3+}. Furthermore, there is one explicit term that was omitted in the earlier work, namely a term that is the complex conjugate of an existing term, which cancels the last term on line $7$ from the final result (also as presented in the erratum). For the momentum space results\cite{Freericks2,Freericks2+}, we find agreement for terms not containing the Peierls substituted hopping---namely, those used to check the final results against the predictions of the Falicov-Kimball model---and our own results. Superficially, the remaining results look quite different, but with some simplification of the final lines of Eqs.~(25,29) in Ref.~\onlinecite{Freericks2} they are brought significantly closer. One must also be careful to remember the ambiguities in the definitions of momentum indices, as the real and imaginary notation employed in this paper must be translated into adding terms indexed by momenta of the opposite sign in order to compare to the earlier work. We have identified the sources of all of these errors in the earlier work, so it does appear that the earlier work did have errors in the terms that were not independently verified. They are all corrected with this work.


\section{Discussion and conclusion}
We have derived a general formula that enables us to evaluate the {\rm $n$th} derivative of a time dependent operator in the Heisenberg representation. We note that this identity can be applied to the full counting statistics problem\cite{Levitov} or to calculating the dynamical algebra of bosons\cite{Richaud}, where one needs to evaluate the derivatives of many operators.

Next, we used these results to evaluate a sequence of spectral moment sum rules for the retarded Green's function and the self-energy in the normal state. These sum rules hold for both nonequilbrium and inhomogeneous cases. The sum rules provide an exact formalism that can benchmark both experimental and computational results. 

For example, use of these sum rules could help decide whether pumping of electrons or phonons can dynamically change the electron-phonon coupling strength. This question is complex. Experimental results clearly show that the kink softens in time-resolved angle-resolved photoemission\cite{Smallwood}, but simulations in a model where the phonon bath has infinite heat capacity and remains fixed in temperature\cite{Kemper} also illustrate a kink softening, yet the zeroth-order moment of the self-energy does not change in the simulation because it is fixed by the parameters in the Hamiltonian. For a bath with finite heat capacity, the net electron-phonon coupling strength will change as the phonon bath is heated since the phonon fluctuations change, which must then change the zeroth moment sum rule for the self-energy.

An interesting future study is the possible extension of these results to the superconducting state. Here, one has both anomalous Greens functions and self-energies, and the structure of the sum rules can change---the retarded Green's function sum rules remain the same, but the self-energies are modified via the modified Dyson equation.
It is possible that sum rules may also help shed light onto nonequilibrium superconductivity\cite{Mitrano} and whether different theories, like the one presented in Ref.~\onlinecite{Sentef} could explain this phenomena. One promising explanation for this phenomenon is non-linearities in the electron-phonon couplings, or ``nonlinear phononics.'' We hope that we will be able to address these issues and expand the sum rules to both the superconducting state and to the case of nonlinear electron-phonon couplings in the future.

\acknowledgments

The early stage of this work was supported by the National Science Foundation under grant
No. EFRI-1433307. The later stage of this work (at Georgetown) was supported by the Department of Energy, Office of Basic Energy
Sciences, Division of Materials Sciences and Engineering under Contract No. DE–SC0019126. 

J.K.F.~was also supported by the McDevitt bequest at Georgetown University. 



\end{document}